\documentclass[aps,prd,eqsecnum,preprint,tightenlines,nofootinbib,showpacs]{revtex4-1}
\usepackage{graphicx,latexsym}

\def\bea{\begin{eqnarray}}
\def\eea{\end{eqnarray}}
\def\be{\begin{equation}}
\def\ee{\end{equation}}

\newcommand{\ob}[1]{\overline{#1}}
\newcommand{\Pminus}{{\cal P}^-}
\newcommand{\pp}{p^{\prime +}}

\begin{document}

\title{Application of the light-front coupled-cluster method \\
to $\phi^4$ theory in two dimensions
}

\author{Blair Elliott}
\author{Sophia S. Chabysheva}
\author{John R. Hiller}
\affiliation{Department of Physics \\
University of Minnesota-Duluth \\
Duluth, Minnesota 55812}

\date{\today}

\begin{abstract}

As a first numerical application of the light-front coupled-cluster (LFCC)
method, we consider the odd-parity massive eigenstate of 
$\phi_{1+1}^4$ theory.  The eigenstate is built as a Fock-state
expansion in light-front quantization, where wave functions
appear as coefficients of the Fock states.  A standard
Fock-space truncation would then yield a finite set of linear
equations for a finite number of wave functions.
The LFCC method replaces Fock-space truncation
with a more sophisticated truncation, one which reduces the eigenvalue
problem to a finite set of nonlinear equations without any
restriction on Fock space.  We compare our results with those obtained
with a Fock-space truncation that yields the same number
of equations.

\end{abstract}

%

\maketitle

\section{Introduction}
\label{sec:Introduction}

The nonperturbative solution of quantum field theories is
considerably more difficult than perturbative calculations.
Various approaches have been developed, with lattice theory~\cite{lattice}
being the most popular.  The Dyson--Schwinger approach~\cite{DSE}
has also had some success.  An alternative is a Hamiltonian
approach based on light-front quantization~\cite{DLCQreviews,Vary,ArbGauge},
which has the advantage of providing boost-invariant wave functions;
such a formulation is much more intuitive, and, of the three, it is
the only one formulated in Minkowski space.

Light-front quantization is done in terms of Dirac's light-front
coordinates~\cite{Dirac}, where $x^+\equiv t+z$ is time and the
corresponding spatial coordinate is $x^-\equiv t-z$.  The transverse
coordinates $x$ and $y$ remain as they were.  The conjugate variables
of light-front energy and momentum are $p^-\equiv E-p_z$ and $p^+\equiv E+p_z$,
respectively.  Again, the transverse components $\vec p_\perp=(p_x,p_y)$
are unchanged.  The mass-shell condition $p^2=m^2$ then implies
$p^-=(m^2+p_\perp^2)/p^+$.  However, here we will be concerned with a
two-dimensional theory, and the transverse components do not enter.

The two-dimensional light-front Hamiltonian eigenvalue problem is~\cite{DLCQreviews},
\be
\Pminus|\psi(P^+)\rangle=\frac{M^2}{P^+}|\psi(P^+)\rangle \;\;
\mbox{and} \;\; 
{\cal P}^+|\psi(P^+)\rangle=P^+|\psi(P^+)\rangle,
\ee
with $\Pminus$ and ${\cal P}^+$ the light-front energy and momentum operators.
The second equation is automatically satisfied by expanding
the eigenstate in Fock states that are themselves eigenstates
of ${\cal P}^+$.  For $n$ bosons with individual momenta $p_i^+$,
we have the total momentum $\sum_i p_i^+=P^+$
and the Fock states $|p_i^+;P^+,n\rangle$.  The eigenstate
expansion is then
\be \label{eq:fockexp}
|\psi(P^+)\rangle=\sum_n (P^+)^{(n-1)/2}
    \int \left(\prod_{i=1}^{n-1} dx_i\right) \psi_n(x_1,...,x_n) |x_iP^+;P^+,n\rangle,
\ee
with $\psi_n$ the $n$-boson wave function.  The factor $(P^+)^{(n-1)/2}$
is explicit in order that $\psi_n$ be independent of $P^+$. 

The Hamiltonian eigenvalue problem now becomes an infinite set of
integral equations for the Fock-state wave functions.  The
standard approximation made is to truncate the expansion to a
finite number of terms, yielding a finite set of equations.
In many theories, such a truncation causes difficulties with
respect to regularization and renormalization, because the
truncation in particle number removes infinite contributions that
would ordinarily cancel against contributions that are retained.
This is the nonperturbative analog of separating a Feynman diagram
into time-ordered diagrams and throwing away diagrams that have
more than some specified number of intermediate particles;
the remaining approximation to the Feynman diagram will
generally be divergent, even if the original covariant diagram
was finite.

One proposed resolution of this difficulty is sector-dependent 
renormalizaton~\cite{Wilson,hb,Karmanov}, where infinities
caused by truncation are absorbed by allowing the bare masses and 
couplings to depend on the Fock sectors involved.  However, this
can lead to ill-defined wave functions~\cite{SecDep}.

The light-front coupled-cluster (LFCC) method~\cite{LFCC} avoids this
by not truncating Fock space.  Instead, the eigenstate
is built as $\sqrt{Z}e^T|\phi\rangle$, where $\sqrt{Z}$ is
a normalizing factor, $|\phi\rangle$ is a valence state with
the smallest possible number of constituents, and $T$ is an
operator that increases particle number.  The truncation 
made is a truncation of $T$; the exponentiation then 
builds all higher Fock states, giving them wave functions
that depend on the valence wave functions and on the functions
in the truncated $T$.

The Hamiltonian eigenvalue problem becomes a valence equation
\be \label{eq:valence}
P_v\ob{\Pminus}|\phi\rangle=\frac{M^2+P_\perp^2}{P^+}|\phi\rangle,
\ee
with $\ob{\Pminus}=e^{-T} \Pminus e^T$ the effective LFCC
Hamiltonian and $P_v$ the projection onto the valence Fock sector.
This is supplemented by an auxiliary equation for $T$
\be \label{eq:aux}
(1-P_v)\ob{\Pminus}|\phi\rangle=0.
\ee
Without truncation, these equations provide an exact
representation of the original eigenvalue problem.
When $T$ is truncated, the projection $1-P_v$ must
also be truncated, to retain only enough equations 
to solve for the functions retained in $T$.  Of course,
the valence eigenvalue problem and the auxiliary equation
must be solved simultaneously.  One immediate advantage
is that the physical mass of the eigenstate appears in
the kinetic-energy terms of the effective Hamiltonian,
without use of sector-dependent renormalization.  The
price to be paid is that the auxiliary equation is
nonlinear in the functions that determine $T$.

When this approach is applied to a simple soluble model~\cite{GreenbergSchweber},
one finds that the simplest truncation of $T$ is sufficient
to generate the exact answer analytically~\cite{LFCC}.
As a first realistic application, where a truncation is an approximation
and where numerical methods are required, we consider here the
odd-parity sector of light-front $\phi^4$ theory in two 
dimensions~\cite{VaryHari}.  The valence state is
the one-particle state, and the $e^T$ operator generates all
the higher odd-particle-number Fock states.  The $T$ operator
is truncated to a single term that creates two additional particles.
By itself, this would only couple the three-particle state, but
the exponentiation of $T$ still generates all the higher odd
states.

The truncated $T$ operator depends on a single function of two
relative momenta.  The auxiliary equation (\ref{eq:aux}) is restricted to
a projection onto the three-particle state, to provide an
equation for this function.  We solve this equation numerically
with use of an expansion in a basis of symmetric multivariate
polynomials~\cite{SymPolys}.  The valence eigenvalue problem
then yields the mass $M$ of the eigenstate.

We limit the calculation to one without symmetry breaking.
The negative mass-squared case, with its wine-bottle potential,
requires careful consideration of the vacuum 
state~\cite{RozowskyThorn,Varyetal}.  An
extension of the LFCC method to handle this has been developed~\cite{LFCCzeromodes},
but we do not apply it here, in order to not complicate the discussion
of this first numerical LFCC application.

Instead, we focus on a comparison of the LFCC calculation
with a Fock-space truncation calculation.  Specifically,
we consider a truncation that yields the same number of
equations and therefore the same computational effort (aside
from the nonlinearity in the LFCC equations).  This is a 
truncation to at most three constituents.  One can then
see explicitly the sector dependence induced by Fock-space truncation
and avoided by the LFCC truncation.

The details of the LFCC analysis are given in Sec.~\ref{sec:analysis},
along with those of the direct Fock-space truncation, for comparison.
Results of numerical solutions are shown and discussed
in Sec.~\ref{sec:results}.  A brief summary is presented in Sec.~\ref{sec:summary}.
The specifics of the simplification of the LFCC equations are left to
Appendix~\ref{sec:reduction}, and Appendix~\ref{sec:methods}
summarizes the numerical methods and illustrates their convergence.

\section{Analysis}
\label{sec:analysis}

The Lagrangian for two-dimensional $\phi^4$ theory is
\be
{\cal L}=\frac12(\partial_\mu\phi)^2-\frac12\mu^2\phi^2-\frac{\lambda}{4!}\phi^4,
\ee
where $\mu$ is the mass of the boson and $\lambda$ is the coupling constant.
The light-front Hamiltonian density is
\be
{\cal H}=\frac12 \mu^2 \phi^2+\frac{\lambda}{4!}\phi^4.
\ee
The mode expansion for the field at zero light-front time is
\be \label{eq:mode}
\phi=\int \frac{dp^+}{\sqrt{4\pi p^+}}
   \left\{ a(p^+)e^{-ip^+x^-/2} + a^\dagger(p^+)e^{ip^+x^-/2}\right\},
\ee
with the modes quantized such that 
\be
[a(p^+),a^\dagger(\pp)]=\delta(p^+-\pp).
\ee
The operator $a^\dagger(p^+)$ creates a boson with momentum $p^+$
and builds the Fock states from the Fock vacuum $|0\rangle$ in the form
\be
|x_iP^+;P^+,n\rangle=\frac{1}{\sqrt{n!}}\prod_{i=1}^n a^\dagger(x_iP^+)|0\rangle.
\ee

\begin{figure}[ht]
\vspace{0.2in}
\begin{center}
\begin{tabular}{c}
\includegraphics[width=15cm]{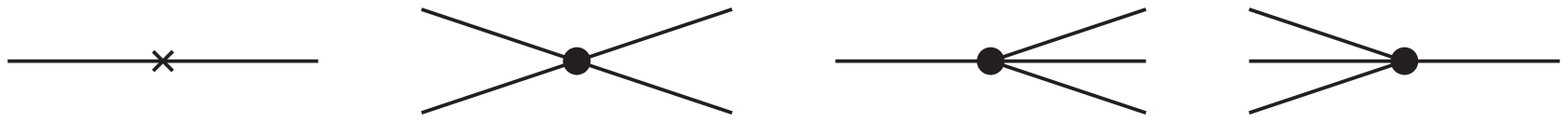} \\
(a) \\ \\
\includegraphics[width=4.5cm]{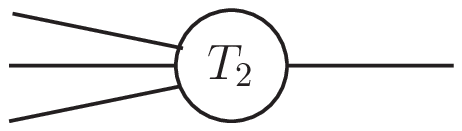} \\
(b)
\end{tabular}
\end{center}
\caption{\label{fig:PT} Graphical representations of (a) the light-front
Hamiltonian $\Pminus$ and (b) the truncated LFCC operator $T_2$.
Lines on the right represent annihilation operators, and those on
the left, creation operators.  The cross is the kinetic energy contribution.
}
\end{figure}

The light-front Hamiltonian for $\phi^4$ theory is 
$\Pminus=\Pminus_{11}+\Pminus_{22}+\Pminus_{13}+\Pminus_{31}$,
with
\bea \label{eq:Pminus11}
\Pminus_{11}&=&\int dp^+ \frac{\mu^2}{p^+} a^\dagger(p^+)a(p^+),  \\
\label{eq:Pminus22}
\Pminus_{22}&=&\frac{\lambda}{4}\int\frac{dp_1^+ dp_2^+}{4\pi\sqrt{p_1^+p_2^+}}
       \int\frac{dp_1^{\prime +}dp_2^{\prime +}}{\sqrt{p_1^{\prime +} p_2^{\prime +}}} 
       \delta(p_1^+ + p_2^+-p_1^{\prime +}-p_2^{\prime +}) \\
 && \rule{2in}{0mm} \times a^\dagger(p_1^+) a^\dagger(p_2^+) a(p_1^{\prime +}) a(p_2^{\prime +}),
   \nonumber \\
\label{eq:Pminus13}
\Pminus_{13}&=&\frac{\lambda}{6}\int \frac{dp_1^+dp_2^+dp_3^+}
                              {4\pi \sqrt{p_1^+p_2^+p_3^+(p_1^++p_2^++p_3^+)}} 
     a^\dagger(p_1^++p_2^++p_3^+)a(p_1^+)a(p_2^+)a(p_3^+), \\
\label{eq:Pminus31}
\Pminus_{31}&=&\frac{\lambda}{6}\int \frac{dp_1^+dp_2^+dp_3^+}
                              {4\pi \sqrt{p_1^+p_2^+p_3^+(p_1^++p_2^++p_3^+)}} 
      a^\dagger(p_1^+)a^\dagger(p_2^+)a^\dagger(p_3^+)a(p_1^++p_2^++p_3^+).
\eea
A graphical representation is given in Fig.~\ref{fig:PT}(a).
The subscripts indicate the number of creation and annihilation operators
in each term; this notation will allow a simple representation of the
terms in the effective Hamiltonian $\ob{\Pminus}$.
Because the terms of $\Pminus$ change particle
number by zero or by two, the eigenstates can be separated 
according to the oddness or evenness of the number of constituents.

We consider the odd case, for which the valence state $|\phi\rangle$ is just the
one-particle state $a^\dagger(P^+)|0\rangle$.  The simplest contribution to
the $T$ operator is
\be \label{eq:T2}
T_2=\int dp_1^+ dp_2^+ dp_3^+ t_2(p_1^+,p_2^+,p_3^+) 
    a^\dagger(p_1^+)a^\dagger(p_2^+)a^\dagger(p_3^+)a(p_1^++p_2^++p_3^+),
\ee
with $t_2$ symmetric in its arguments, and we truncate $T$ to this,
as represented in Fig.~\ref{fig:PT}(b).  The 
projection $1-P_v$ for the auxiliary equation (\ref{eq:aux}) is then just
a projection onto the three-particle state $a^\dagger(p_1^+)a^\dagger(p_2^+)a^\dagger(p_3^+)|0\rangle$.
The effective Hamiltonian $\ob{\Pminus}$ can be constructed from the
Baker--Hausdorff expansion $e^{-T}\Pminus e^T=\Pminus+[\Pminus,T]+\frac12[[\Pminus,T],T]+\cdots$.
The series can be terminated, without additional approximation, when the net increase
in particle number is greater than what is needed for the truncated auxiliary equation.
However, this generates many terms that do not actually contribute to the
valence equation or to the auxiliary equation.  Therefore, a more efficient approach
for the construction of these equations is to compute explicitly the matrix elements
of $\ob{\Pminus}$ that enter into the projections onto the valence one-particle
sector and the three-particle sector.

\begin{figure}[ht]
\vspace{0.2in}
\begin{center}
\begin{tabular}{c}
\includegraphics[width=7.5cm]{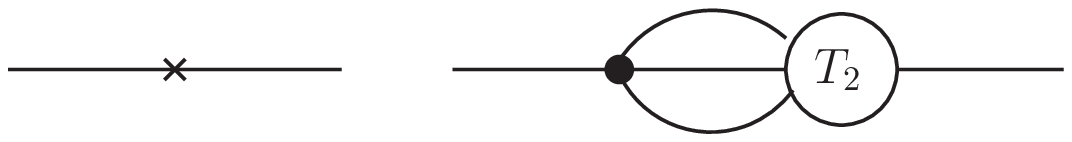} \\
(a) \\ \\
\includegraphics[width=16cm]{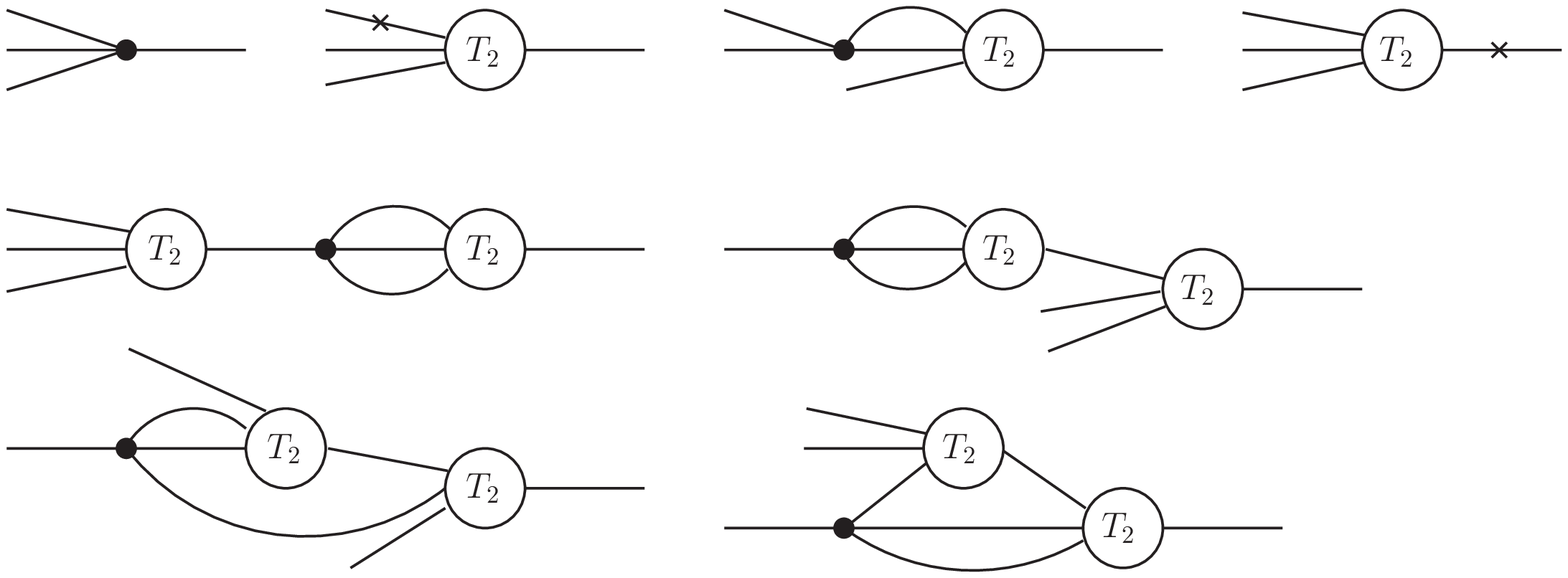} \\
(b)
\end{tabular}
\end{center}
\caption{\label{fig:LFCCeqns} Graphical representations of the (a) valence 
and (b) auxiliary equations, given in Eqs.~(\ref{eq:valenceprojected}) and
(\ref{eq:auxprojected}) of the text.
}
\end{figure}

The valence equation (\ref{eq:valence}) has the following contributions:
\be \label{eq:valenceprojected}
\langle 0|a(Q^+)\left(\Pminus_{11}+\Pminus_{13}T_2\right)a^\dagger(P^+)|0\rangle
               =\frac{M^2}{P^+}\delta(Q^+-P^+).
\ee
The auxiliary equation (\ref{eq:aux}) becomes
\be \label{eq:auxprojected}
\langle 0|a(q_1^+)a(q_2^+)a(q_3^+)\left(\Pminus_{31}+(\Pminus_{11}+\Pminus_{22})T_2-T_2\Pminus_{11}
          -T_2\Pminus_{13}T_2+\frac12\Pminus_{13}T_2^2\right)a^\dagger(P^+)|0\rangle=0.
\ee
Graphical representations are given in Fig.~\ref{fig:LFCCeqns}.
The reduction of these equations, including the evaluation of the individual
matrix elements, is carried out in Appendix~\ref{sec:reduction}.
The valence equation is reduced to
\be \label{eq:LFCCvalence}
1+g\int\frac{dx_1 dx_2}{\sqrt{x_1 x_2 x_3}}\tilde t_2(x_1,x_2,x_3)=M^2/\mu^2,
\ee
where the $x_i=p_i^+/P^+$ are longitudinal momentum fractions, with $x_3=1-x_1-x_2$,
$g$ is a dimensionless coupling constant, defined by
\be \label{eq:g}
g=\frac{\lambda}{4\pi\mu^2},
\ee
and $\tilde t_2$ is a rescaled function of longitudinal momentum fractions,
\be \label{eq:tildet2}
\tilde t_2(x_1,x_2,x_3)=P^+t_2(x_1P^+,x_2P^+,x_3P^+).
\ee
This leads to the definition of a dimensionless mass shift $\Delta$
\be \label{eq:Delta}
\Delta\equiv g\int\frac{dx_1 dx_2}{\sqrt{x_1 x_2 x_3}}\tilde t_2(x_1,x_2,x_3),
\ee
such that $M^2=(1+\Delta)\mu^2$.
Thus the valence equation determines the mass $M$, given the reduced
$T$ function $\tilde t_2$.

The final form of the auxiliary equation is~\cite{Elliott}
\bea \label{eq:LFCCaux}
\lefteqn{\frac16\frac{g}{\sqrt{y_1 y_2 y_3}}
             +(1+\Delta)\left(\frac{1}{y_1}+\frac{1}{y_2}+\frac{1}{y_3}-1\right)
                                    \tilde t_2(y_1,y_2,y_3)}&& \\
   &&  +\frac{g}{2}\left[\int_0^{1-y_1}dx_1
             \frac{\tilde t_2(y_1,x_1,1-y_1-x_1)}{\sqrt{x_1 y_2 y_3 (1-y_1-x_1)}} 
               + (y_1 \leftrightarrow y_2) + (y_1 \leftrightarrow y_3)\right] \nonumber \\
    &&  -\frac{\Delta}{2} \left(\frac{1}{y_1}+\frac{1}{y_2}+\frac{1}{y_3}\right) \tilde t_2(y_1,y_2,y_3)
              \nonumber \\
    && +\frac{3g}{2}\left\{\int_{y_1/(1-y_2)}^1 d\alpha_1 \int_0^{1-\alpha_1} d\alpha_2 
       \frac{\tilde t_2(y_1/\alpha_1,y_2,1-y_1/\alpha_1-y_2) \tilde t_2(\alpha_1,\alpha_2,\alpha_3)}
       {\sqrt{\alpha_1 \alpha_2 \alpha_3 y_3 (\alpha_1-y_1-\alpha_1 y_2)}}\right.      \nonumber \\
    && \rule{2.2in}{0mm} \left.  +(y_1\leftrightarrow y_2)+(y_1\leftrightarrow y_3)
                                               \rule{0mm}{0.25in} \right\} \nonumber \\
    && +\frac{3g}{2}\left\{ \left[ \int_{y_1+y_2}^1 d\alpha_1 \int_0^{1-\alpha_1} d\alpha_2
        \frac{\tilde t_2(y_1/\alpha_1,y_2/\alpha_1,1-(y_1+y_2)/\alpha_1)\tilde t_2(\alpha_1,\alpha_2,\alpha_3)}
        {\alpha_1 \sqrt{\alpha_2 \alpha_3 y_3 (\alpha_1-y_1-y_2)}} \right.  \right.   \nonumber \\
    && \rule{2.5in}{0mm} \left.  + (y_2 \leftrightarrow y_3)
                                                   \rule{0mm}{0.25in}\right]  \nonumber \\
     && \rule{1in}{0mm}  \left. +(y_1\leftrightarrow y_2)+(y_1\leftrightarrow y_3)
                                        \rule{0mm}{0.25in}\right\}=0, \nonumber
\eea
with $y_i=q_i^+/P^+$. 
It is this form of the auxiliary equation that we solve numerically, to obtain
$\tilde t_2$, as discussed in Appendix~\ref{sec:methods}.

\begin{figure}[ht]
\vspace{0.2in}
\begin{center}
\begin{tabular}{c}
\includegraphics[width=8cm]{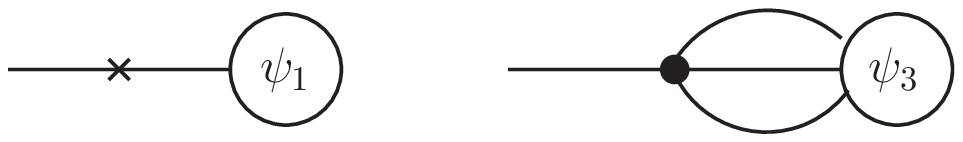} \\
(a) \\ \\
\includegraphics[width=12cm]{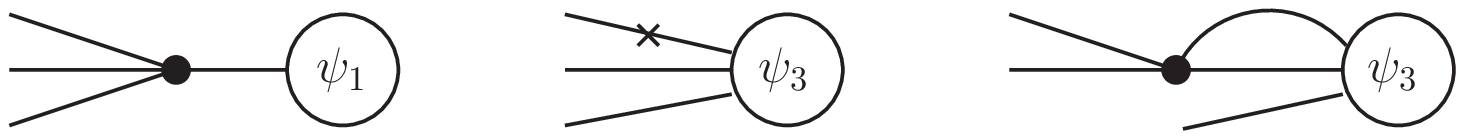} \\
(b)
\end{tabular}
\end{center}
\caption{\label{fig:truncated} Graphical representations of the (a) one-body
and (b) three-body equations in the case of Fock-space truncation,
given in Eqs.~(\ref{eq:tildepsi1}) and (\ref{eq:tildepsi3}) of the text.
}
\end{figure}

For comparison, we consider the Fock-space truncation approach, with the number
of constituents limited to three, so that the resulting equations have the
same number of independent variables as the LFCC equations derived above.
The eigenstate is then approximated by 
\be
|\psi(P^+)\rangle=\psi_0 a^\dagger(P^+)|0\rangle 
      +P^+\int dx_1 dx_2 \psi_3(x_1,x_2,x_3)\frac{1}{\sqrt{3!}}
         a^\dagger(x_1P^+)a^\dagger(x_2P^+)a^\dagger(x_3P^+)|0\rangle.
\ee
The action of the light-front Hamiltonian $\Pminus$
then yields the following coupled system of integral equations:
\bea
\lefteqn{\mu^2\psi_1
  +\frac{\lambda}{6}\int\frac{dx_1 dx_2}{4\pi\sqrt{x_1 x_2 x_3}}\psi_3(x_1,x_2,x_3)=M^2\psi_1,}&& \\
&& \mu^2\left(\frac{1}{y_1}+\frac{1}{y_2}+\frac{1}{y_3}\right)\psi_3(y_1,y_2,y_3)
  +\lambda\frac{\psi_1}{4\pi\sqrt{y_1y_2y_3}} \\
     &&  +\frac12\frac{\lambda}{4\pi}\left[ \int_0^{1-y_1} dx_1
             \frac{\psi_3(x_1,y_1,1-y_1-x_1)}{\sqrt{x_1(1-y_1-x_1)y_2 y_3}} 
               + (y_1 \leftrightarrow y_2) + (y_1 \leftrightarrow y_3)\rule{0mm}{0.3in}\right]
                   =M^2\psi_3(y_1,y_2,y_3).   \nonumber
\eea
If we define $\tilde\psi_3=\psi_3/(6\psi_1)$, these reduce to forms
directly comparable to the LFCC equations
\bea \label{eq:tildepsi1}
\lefteqn{1+g\int\frac{dx_1 dx_2}{\sqrt{x_1 x_2 x_3}}\tilde\psi_3(x_1,x_2,x_3)=M^2/\mu^2,} && \\
\label{eq:tildepsi3}
&& \frac16\frac{g}{\sqrt{y_1y_2y_3}}
   +\left(\frac{1}{y_1}+\frac{1}{y_2}+\frac{1}{y_3}-\frac{M^2}{\mu^2}\right)\tilde\psi_3(y_1,y_2,y_3) \\
     &&  +\frac{g}{2}\left[ \int_0^{1-y_1} dx_1
             \frac{\tilde\psi_3(x_1,y_1,1-y_1-x_1)}{\sqrt{x_1(1-y_1-x_1)y_2 y_3}} 
               + (y_1 \leftrightarrow y_2) + (y_1 \leftrightarrow y_3)\rule{0mm}{0.3in}\right]
                   =0.   \nonumber
\eea
Graphical representations are given in Fig.~\ref{fig:truncated}.

The first Fock-state equation (\ref{eq:tildepsi1}) is the same as the LFCC valence
equation (\ref{eq:LFCCvalence}); it relates
the physical mass to the bare mass through the self-energy correction due to the
three-particle intermediate state.  However, the second equation differs in
several respects.  The inhomogeneous term, from the coupling to the one-particle
state, and the two-two scattering term, the last term in (\ref{eq:tildepsi3}),
remain the same.  The eigenvalue term that appears in the equation for
$\tilde\psi_3$ is also seen to be present in the LFCC auxiliary equation,
with use of the valence relation, $1+\Delta=M^2/\mu^2$, but the kinetic energy
contributions for the three individual constituents are not the same.  In
the $\tilde\psi_3$ equation, they enter only as $\mu^2/y_i$, whereas in
the LFCC auxiliary equation, they are $M^2/y_i$, again with $1+\Delta=M^2/\mu^2$.
The sector-dependent approach~\cite{Wilson,hb,Karmanov,SecDep} would rectify this
by making $\mu$ sector dependent and equal to $M$ in the highest Fock sector;
this compensates for the lack of a self-energy correction in a sector
for which there can be no additional particles to generate such a loop.
The LFCC approach automatically inserts the correct mass without using
a sector-dependent bare mass.

The LFCC auxiliary equation also contains several terms that do not appear
in the second Fock-state equation.  The fourth term is the nonperturbative
analog of the wave-function renormalization counterterm, which is
a subtraction from the loop contributions represented by the fifth
and sixth terms.  These last two terms are a partial resummation
of the high-order loops generated by Fock sectors beyond the
three-particle sector.  The resummation is partial, because
the truncated $T$ operator is an approximation.  

The price to be paid for these additions is the nonlinearity
of the LFCC auxiliary equation.  However, this does not present
any particular difficulty for its solution.  Our numerical
methods are described in Appendix~\ref{sec:methods}.  They rely
on an expansion of $\tilde t_2$ in a basis of fully symmetric
polynomials~\cite{SymPolys}, which converts the auxiliary
equation to a system of nonlinear algebraic equations.  The
results obtained in this way are presented and discussed in
the next section.  We also solve the Fock-state wave-function
equations in the same way, for numerical comparison.

\section{Results}
\label{sec:results}

\begin{figure}[ht]
\vspace{0.2in}
\centerline{\includegraphics[width=12cm]{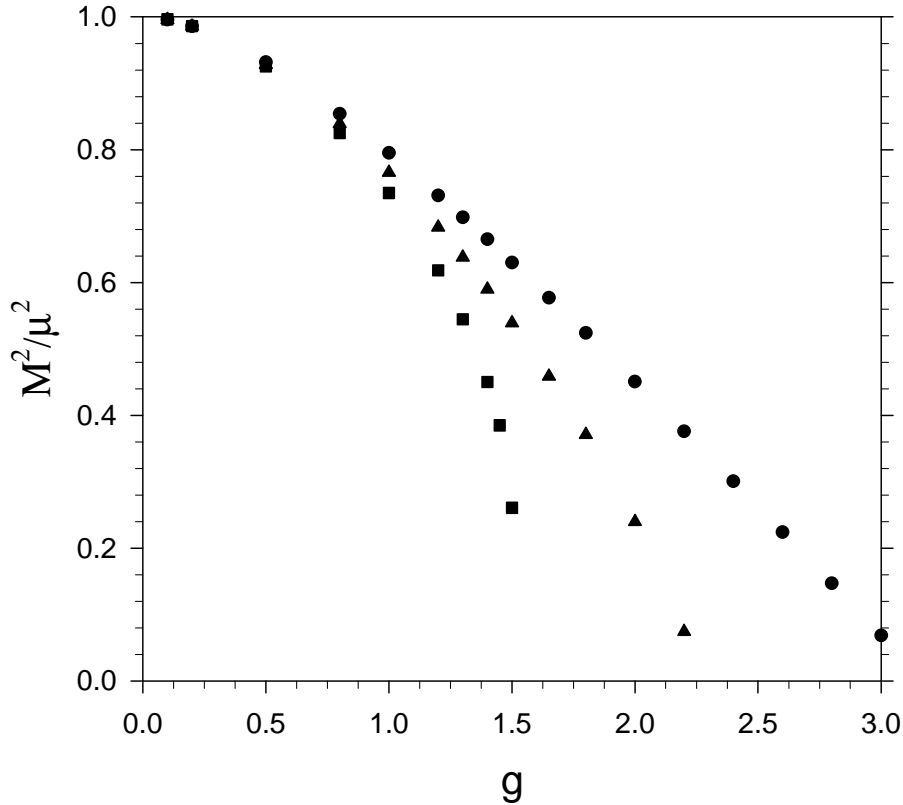}}
\caption{\label{fig:M2vsg} Mass-squared ratios $M^2/\mu^2$ versus
dimensionless coupling strength $g$ for the LFCC approximation (squares),
the Fock-space truncation (circles), and the Fock-space truncation
with sector-dependent masses (triangles).
}
\end{figure}

The converged results for the mass-squared eigenvalues are
shown in Fig.~\ref{fig:M2vsg}.  This figure compares results
for the LFCC approximation with those obtained with a
Fock-space truncation.  In the latter case, two different
approximations are considered, with and without sector-dependent
bare masses.  When sector-dependent masses are used, the
leading $\mu^2$ in Eq.~(\ref{eq:tildepsi3}) is replaced by
$M^2$; this compensates for the exclusion of
any self-energy corrections in the top Fock sector
by the Fock-space truncation.

There are significant
differences in behavior among the three cases, with
the LFCC result decreasing most rapidly with increasing
coupling.  The sector-dependent case proves to be 
intermediate, as expected; one of the improvements that
the LFCC method offers is to place sector-dependent
masses automatically.  The remainder of the difference,
which neither Fock-space truncation can include, is
the resummation of contributions from all higher 
Fock states.  

Also, unlike the two Fock-space truncation
cases, which continue to zero and below with increased
coupling, the LFCC eigenmass becomes complex at approximately 
$g=1.5$ instead of continuing toward zero.  Since the
approach to zero is associated with spontaneous symmetry 
breaking, we expect that zero modes will be required to
do a proper analysis~\cite{LFCCzeromodes}.

\begin{figure}[ht]
\vspace{0.2in}
\centerline{
\includegraphics[width=12cm]{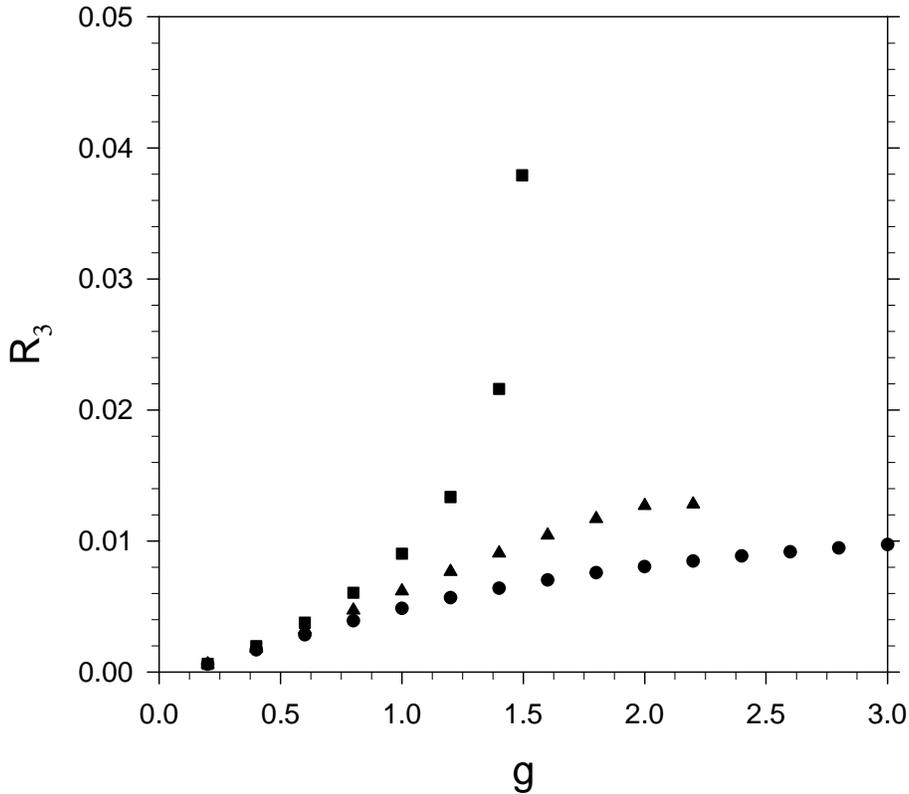}}
\caption{\label{fig:relprob} Same as Fig.~\ref{fig:M2vsg},
but for the relative probability $R_3$.
}
\end{figure}

Another signal that the LFCC approximation provides more
information is in the probabilities for higher
Fock states.  In Fig.~\ref{fig:relprob}, we plot the
probability for the three-body Fock state, relative to
the one-body state, as a function of the coupling.
In general, the relative probability for the $n$-body
state is zero for $n$ even and given as follows for
$n$ odd:
\be
R_n=\frac{1}{Z}\int\left(\prod_{i=1}^{n-1} dx_i\right) |\psi_n(x_i)|^2,
\ee
where $Z$ is the probability for the one-body state and $\psi_n$ is
the wave function for the $n$-body state, as in (\ref{eq:fockexp}).  In the LFCC method,
the chosen truncation (\ref{eq:T2}) of $T$ yields these wave functions as
\be
\psi_n(x_i)=\int dP^{\prime +} \frac{1}{\sqrt{n!}}
      \langle 0|\prod_{i=1}^n a(x_iP^{\prime +})
         \frac{(P^+)^{(n-1)/2}\sqrt{Z}}{((n-1)/2)!}T_2^{(n-1)/2}a^\dagger(P^+)|0\rangle.
\ee
The integration over $P^{\prime +}$ eliminates the momentum conserving
delta function from the projection.  The plots in Fig.~\ref{fig:relprob} show
that the relative probability $R_3$ increases rapidly in the LFCC approximation
as the coupling approaches the value at which the mass value becomes complex.
The Fock-space truncation results remain slowly varying, even as the 
coupling approaches values at which the mass value becomes zero.

\begin{figure}[ht]
\vspace{0.2in}
\begin{center}
\begin{tabular}{ccc}
\includegraphics[width=5cm]{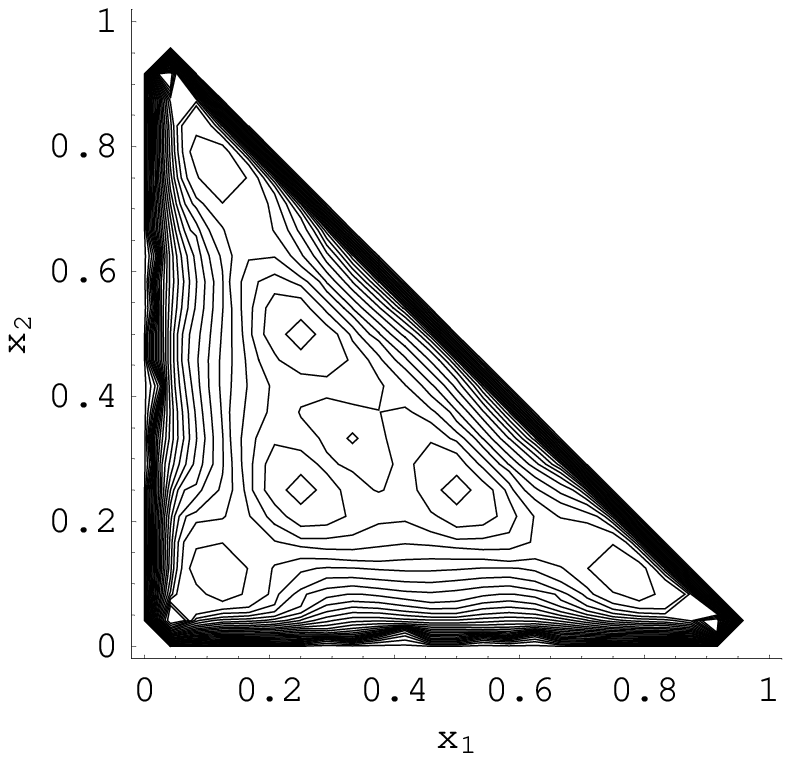} &
\includegraphics[width=5cm]{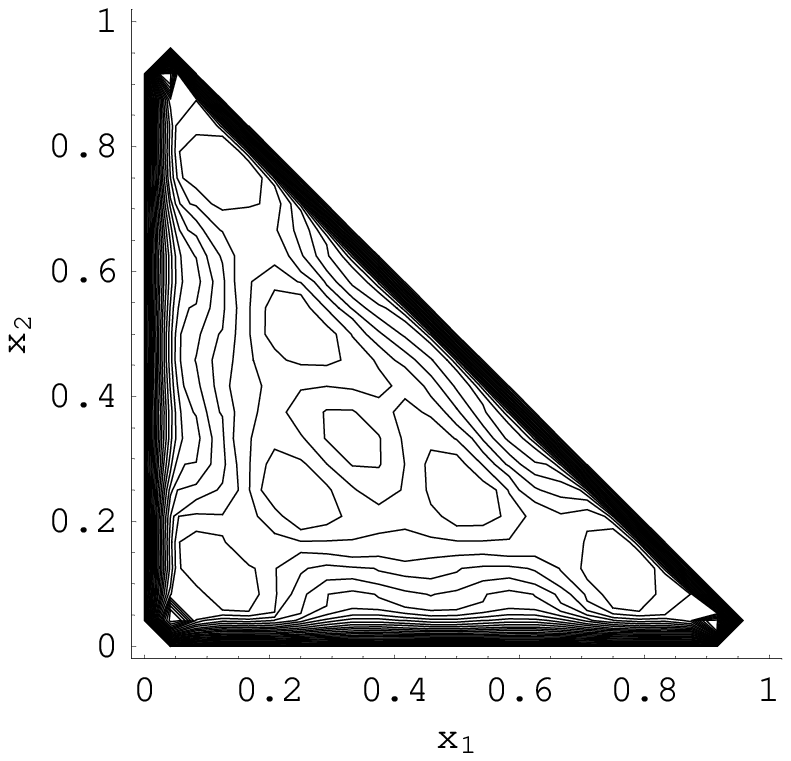} &
\includegraphics[width=5cm]{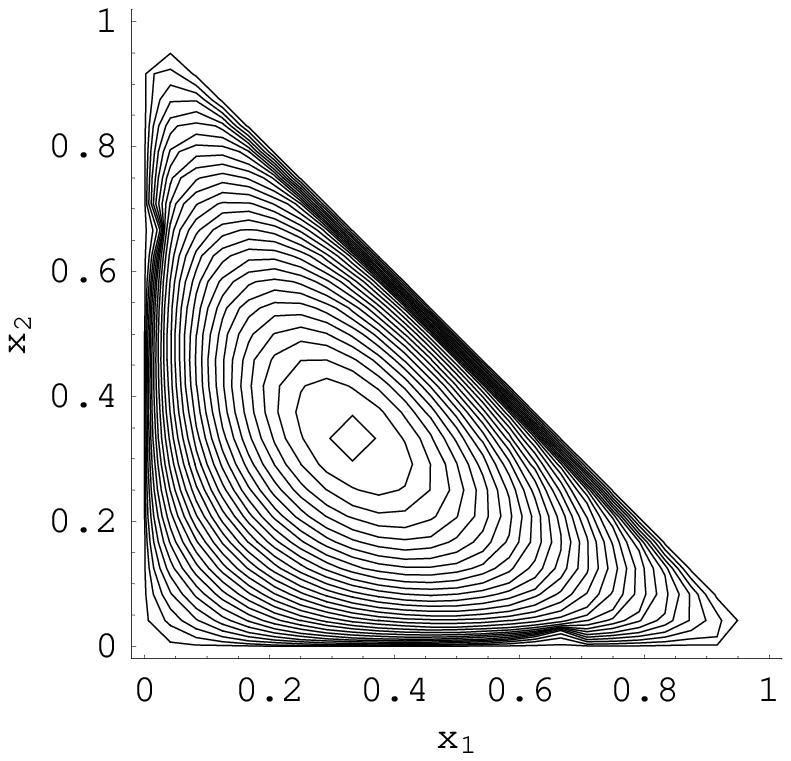} \\
(a) & (b) & (c) 
\end{tabular}
\end{center}
\caption{\label{fig:contour} Contour plots of the three-body
wave function for (a) the LFCC approximation,
(b) the Fock-space truncation, and (c) the Fock-space truncation
without two-two scattering.  The dimensionless coupling strength
was $g=1$ in all three cases.
}
\end{figure}

Figure~\ref{fig:contour} provides a comparison of
three-body wave functions.  For the LFCC
case, what is actually plotted is the $T$ function $\tilde t_2$,
which also determines the higher Fock-state wave functions.
The main qualitative difference between cases is associated
with the two-two scattering process; when it is left out,
as in Fig.~\ref{fig:contour}(c), the structure of the three-body
wave function is much simpler.

\section{Summary}
\label{sec:summary}

We have illustrated the use of the LFCC method~\cite{LFCC} in
an application to a model theory that is simple enough to 
make the method plain, yet complex enough to require
numerical techniques.  This goes beyond the previous
illustration~\cite{LFCC} that used a soluble model~\cite{GreenbergSchweber}.
In particular, a comparison with a Fock-space truncation
approach shows explicitly that the LFCC method introduces
the physical mass for kinetic energy terms without use
of a sector-dependent parameterization~\cite{SecDep},
as is seen in the comparison of Eqs.~(\ref{eq:auxprojected})
and (\ref{eq:tildepsi3}).  In the former, all kinetic
terms contain $1+\Delta\equiv M^2/\mu^2$ and in the
latter, the three-body terms contain only $1=\mu^2/\mu^2$.
Thus, if the physical mass is to be restored in the highest
sector of a truncated Fock space, where no self-energy
loops can occur, the bare mass must be sector dependent 
and set equal to the physical mass only in that highest 
sector.  The LFCC method arranges the mass automatically,
without use of sector-dependent bare masses.

A comparison of numerical results is given in Figs.~\ref{fig:M2vsg},
\ref{fig:relprob}, 
and \ref{fig:contour}.  The calculations done with Fock-space 
truncation clearly yield results which differ those of the
LFCC calculation.  This is to be expected, because the LFCC
method includes effects of higher Fock states.  However, the
three-body wave functions are qualitatively similar, with
the dominant effect being the inclusion of two-two scattering.

Future work along these lines could include various extensions
of the $T$ operator, which could be used to study convergence
of the LFCC results as more terms are added.  Also, the process
of symmetry breaking is of particular interest, for both positive 
and negative $\mu^2$.
A partial analysis can be done by comparing odd and even
eigenstates, by looking for degeneracy in the lowest 
states~\cite{RozowskyThorn,Varyetal}; 
this will require consideration of valence
states with two particles and at least one additional term
in $T$.  A full analysis is best done with inclusion of
modes of zero longitudinal momentum; for this some preliminary
work has already been done~\cite{LFCCzeromodes}.

\acknowledgments
This work was supported in part by the Department of Energy
through Contract No.\ DE-FG02-98ER41087
and by the Minnesota Supercomputing Institute through
grants of computing time.
We thank J. Vary for comments on preliminary results.

\appendix

\section{Reduction of equations}  \label{sec:reduction}

In order to reduce the valence and auxiliary equations
(\ref{eq:valenceprojected}) and (\ref{eq:auxprojected})
to more usable forms, we must first evaluate the
matrix elements of products of $\Pminus_{ij}$ and
powers of $T_2$.  The expressions for these operators
are given in Eqs.~(\ref{eq:Pminus11})-(\ref{eq:Pminus22})
and Eq.~(\ref{eq:T2}). For the valence equation, we need
\be 
\langle 0|a(Q^+)\Pminus_{11}a^\dagger(P^+)|0\rangle
    =\frac{\mu^2}{P^+}\delta(Q^+-P^+)
\ee
and
\be 
\langle 0|a(Q^+)\Pminus_{13}T_2 a^\dagger(P^+)|0\rangle
   =\delta(Q^+-P^+)\frac{\lambda}{4\pi}
     \int \frac{dp_1^+ dp_2^+ dp_3^+ t_2(p_1^+, p_2^+, p_3^+)}
               {\sqrt{p_1^+ p_2^+ p_3^+(p_1^++p_2^++p_3^+)}}
       \delta(P^+-p_1^+-p_2^+-p_3^+),
\ee
where we have made of use of the symmetry of $t_2$ to reduce $3!$ 
contraction terms to only one.
For the auxiliary equation, we require
\be 
\langle 0|a(q_1^+)a(q_2^+)a(q_3^+)\Pminus_{31}a^\dagger(P^+)|0\rangle=
\frac{\lambda}{4\pi}\frac{\delta(P^+-q_1^+-q_2^+-q_3^+)}
                         {\sqrt{q_1^+ q_2^+ q_3^+(q_1^++q_2^++q_3^+)}},
\ee
\bea
\lefteqn{\langle 0|a(q_1^+)a(q_2^+)a(q_3^+)(\Pminus_{11}T_2-T_2\Pminus_{11})a^\dagger(P^+)|0\rangle}&& \\
&& \rule{0.5in}{0mm}
=6\delta(P^+-q_1^+-q_2^+-q_3^+)\left(\frac{\mu^2}{q_1^+}+\frac{\mu^2}{q_2^+}+\frac{\mu^2}{q_3^+}
      -\frac{\mu^2}{P^+}\right) t_2(q_1^+,q_2^+,q_3^+), \nonumber
\eea
\bea
\lefteqn{\langle 0|a(q_1^+)a(q_2^+)a(q_3^+)\Pminus_{22} T_2 a^\dagger(P^+)|0\rangle} && \\
&& \rule{0.5in}{0mm}
=3\delta(P^+-q_1^+-q_2^+-q_3^+)\frac{\lambda}{4\pi}\int_0^{1-q_1^+}
   \frac{dp_2^+ t_2(q_1^+,p_2^+,1-q_1^+-p_2^+)}{\sqrt{p_2^+(1-q_1^+-p_2^+)q_2^+ q_3^+}} \nonumber \\
 && \rule{1in}{0mm}  + (q_1^+\leftrightarrow q_2^+)+ (q_1^+\leftrightarrow q_3^+), \nonumber
\eea
\bea
\lefteqn{\langle 0|a(q_1^+)a(q_2^+)a(q_3^+)T_2\Pminus_{13}T_2 a^\dagger(P^+)|0\rangle} && \\
&& \rule{0.5in}{0mm}
=6\delta(P^+-q_1^+-q_2^+-q_3^+)\frac{\lambda}{4\pi}t_2(q_1^+,q_2^+,q_3^+)\nonumber \\
 && \rule{1in}{0mm} \times
\int \frac{dp_1^+ dp_2^+ dp_3^+}{\sqrt{p_1^+ p_2^+ p_3^+ P^+}} 
   t_2(p_1^+, p_2^+, p_3^+) \delta(P^+-p_1^+-p_2^+-p_3^+), \nonumber
\eea
and
\bea
\lefteqn{\langle 0|a(q_1^+)a(q_2^+)a(q_3^+)\Pminus_{13}T_2^2 a^\dagger(P^+)|0\rangle} && \\
&& \rule{0.5in}{0mm}
= 3\delta(P^+-q_1^+-q_2^+-q_3^+)\frac{\lambda}{4\pi}
\int\frac{dp_1^+ dp_2^+ dp_3^+ \delta(q_1^+-p_1^+ -p_2^+ -p_3^+)}
            {\sqrt{p_1^+ p_2^+ p_3^+ q_1^+}} \nonumber \\
 && \rule{1in}{0mm} \times
   \left\{ 3 t_2(p_1^+,q_2^+,q_3^+)t_2(p_1+q_2^++q_3^+,p_2^+,p_3^+)
          + t_2(p_1^+,p_2^+,p_3^+) t_2(q_1^+,q_2^+,q_3^+) \right. \nonumber \\
 && \rule{1.7in}{0mm} \left.
          +3\left[t_2(p_1^+,q_2^+,p_3^+) t_2(p_1^++p_3^++q_2^+,p_2^+,q_3^+)
            + (q_2^+\leftrightarrow q_3^+)\right]\right\} \nonumber \\
 && \rule{0.7in}{0mm}
            +(q_1^+ \leftrightarrow q_2^+) + (q_1^+ \leftrightarrow q_3^+). \nonumber
\eea

For the valence equation, substitution of the first two matrix elements yields
\be \label{eq:intermediatevalence}
\frac{\mu^2}{P^+}+\frac{\lambda}{4\pi}\int
  \frac{dp_1^+ dp_2^+ dp_3^+}{\sqrt{p_1^+ p_2^+ p_3^+}}t_2(p_1^+,p_2^+,p_3^+)
  \delta(P^+-p_1^+-p_2^+-p_3^+)=\frac{M^2}{P^+}.
\ee
In terms of the dimensionless coupling constant $g$, defined in Eq.~(\ref{eq:g}), and
the reduced $T$ function $\tilde t_2$, defined in Eq.~(\ref{eq:tildet2}), the valence
equation (\ref{eq:intermediatevalence}) takes the form given in Eq.~(\ref{eq:LFCCvalence}).

For the auxiliary equation (\ref{eq:auxprojected}), evaluation of the matrix elements provides
\bea
\lefteqn{\left\{\frac16\frac{\lambda}{4\pi}\frac{1}{\sqrt{q_1^+ q_2^+ q_3^+ P^+}}
             +\left(\frac{\mu^2}{q_1^+}+\frac{\mu^2}{q_2^+}+\frac{\mu^2}{q_3^+}
                           -\frac{\mu^2}{P^+}\right)t_2(q_1^+,q_2^+,q_3^+)\right.}&& \\
   &&  +\frac12\frac{\lambda}{4\pi}\left[
        \frac{1}{\sqrt{q_2^+ q_3^+}}\int_0^{q_2^++q_3^+}
             \frac{dp_1^+}{\sqrt{p_1^+(q_2^++q_3^+-p_1^+)}}t_2(p_1^+,q_2^++q_3^+-p_1^+,q_1^+) \right. \nonumber \\
    && \left. \rule{3in}{0mm}
               + (q_1^+ \leftrightarrow q_2^+) + (q_1^+ \leftrightarrow q_3^+)\rule{0mm}{0.3in}\right] \nonumber \\
    && -\frac{\lambda}{4\pi}t_2(q_1^+,q_2^+,q_3^+)
           \int\frac{dp_1^+ dp_2^+ dp_3^+}{\sqrt{p_1^+ p_2^+ p_3^+ P^+}}
              \delta(P^+-p_1^+-p_2^+-p_3^+)t_2(p_1^+,p_2^+,p_3^+) \nonumber \\
    &&  +\frac12\frac{\lambda}{4\pi}\left[
             \int\frac{dp_1^+ dp_2^+ dp_3^+}{\sqrt{p_1^+ p_2^+ p_3^+ q_1^+}}\delta(q_1^+-p_1^+-p_2^+-p_3^+)
                \left\{t_2(p_1^+,p_2^+,p_3^+)t_2(q_1^+,q_2^+,q_3^+) \right. \right. \nonumber \\
     &&  \rule{1in}{0mm}
                   +3 t_2(p_1^+,q_2^+,q_3^+)t_2(p_1^++q_2^++q_3^+,p_2^+,p_3^+) \nonumber \\
      && \left.   \rule{1in}{0mm}     +3\left[t_2(p_1^+,q_2^+,p_3^+)t_2(p_1^++p_3^++q_2^+,p_2^+,q_3^+)
                                       +(q_2^+\leftrightarrow q_3^+)\right]\right\} \nonumber \\
      &&   \left.  \left.   \rule{0.5in}{0mm}
           +(q_1^+\leftrightarrow q_2^+)+(q_1^+\leftrightarrow q_3^+)\rule{0mm}{0.265in}\right]\right\}
                     \delta(P^+-q_1^+-q_2^+-q_3^+)=0. \nonumber
\eea
We define momentum fractions $x_i=p_i^+/P^+$ and $y_i=q_i^+/P^+$ and obtain
\bea
\lefteqn{\frac16\frac{g}{\sqrt{y_1 y_2 y_3}}
             +\left(\frac{1}{y_1}+\frac{1}{y_2}+\frac{1}{y_3}-1\right)\tilde t_2(y_1,y_2,y_3)}&& \\
   &&  +\frac{g}{2}\left[
        \frac{1}{\sqrt{y_2 y_3}}\int_0^{1-y_1}dx_1
             \frac{\tilde t_2(x_1,1-y_1-x_1,y_1)}{\sqrt{x_1(1-y_1-x_1)}} 
               + (y_1 \leftrightarrow y_2) + (y_1 \leftrightarrow y_3)\right] \nonumber \\
    && -g \tilde t_2(y_1,y_2,y_3)
           \int\frac{dx_1 dx_2}{\sqrt{x_1 x_2 x_3}} \tilde t_2(x_1,x_2,x_3) \nonumber \\
    &&  +\frac{g}{2} \tilde t_2(y_1,y_2,y_3)\left(\frac{1}{y_1}+\frac{1}{y_2}+\frac{1}{y_3}\right)
             \int\frac{dx_1 dx_2}{\sqrt{x_1 x_2 x_3}}\tilde t_2(x_1,x_2,x_3) \nonumber \\
    && +\frac{3g}{2}\left\{\int\frac{dx_1 dx_2}{\sqrt{x_1 x_2 x_3}}\frac{1}{x_1y_1+1-y_1}
        \tilde t_2\left(\frac{x_1 y_1}{x_1y_1+1-y_1},\frac{y_2}{x_1y_1+1-y_1},\frac{y_3}{x_1y_1+1-y_1}\right) 
                                  \right.      \nonumber \\
    && \rule{2.2in}{0mm} \left. \times
        \tilde t_2(x_1y_1+1-y_1,x_2y_1,x_3y_1) +(y_1\leftrightarrow y_2)+(y_1\leftrightarrow y_3)
                                               \rule{0mm}{0.25in} \right\} \nonumber \\
    && +\frac{3g}{2}\left\{\int \frac{dx_1 dx_2}{\sqrt{x_1 x_2 x_3}}
          \left[\frac{1}{(1-x_2)y_1+y_2}
        \tilde t_2\left(\frac{x_1 y_1}{(1-x_2)y_1+y_2},\frac{y_2}{(1-x_2)y_1+y_2},\frac{x_3 y_1}{(1-x_2)y_1+y_2}\right)
                                 \right.  \right.     \nonumber \\
    && \rule{2.5in}{0mm} \left. \times
                    \tilde t_2((1-x_2)y_1+y_2,x_2y_1,y_3) + (y_2 \leftrightarrow y_3)
                                                   \rule{0mm}{0.25in}\right]  \nonumber \\
     && \rule{1in}{0mm}  \left. +(y_1\leftrightarrow y_2)+(y_1\leftrightarrow y_3)
                                        \rule{0mm}{0.25in}\right\}=0, \nonumber
\eea
where $\tilde t_2$ is the rescaled function defined in Eq.~(\ref{eq:tildet2}).
To further simplify the auxiliary equation, we introduce changes of variable for
the next-to-last and last sets of terms.  For the next-to-last, we define
$\alpha_1=x_1 y_1+1-y_1$, $\alpha_2=x_2y_1$, and $\alpha_3=1-\alpha_2-\alpha_3=x_3 y_1$.
For the last, we define $\alpha_1=x_1 y_1/((1-x_2)y_1+y_2)$, $\alpha_2=x_3 y_1/((1-x_2)y_1+y_2)$,
and $\alpha_3=1-\alpha_1-\alpha_2=y_3/((1-x_2)y_1+y_2)$.  We also take advantage
of the symmetry of $\tilde t_2$, to interchange the order of its arguments, and
identify the self-energy correction $\Delta$ to the mass, as defined in
Eq.~(\ref{eq:Delta}).  These yield the final expression given in Eq.~(\ref{eq:LFCCaux}).

\section{Numerical methods}  \label{sec:methods}

We solve the integral equation (\ref{eq:LFCCaux}) by expanding
the function $\tilde t_2$ in a basis of fully symmetric
polynomials~\cite{SymPolys}
\be
\tilde t_2(x_1,x_2,x_3)=\sqrt{x_1 x_2 x_3}\sum_{n,i}^{n=N} a_{ni}P_{ni}(x_1,x_2).
\ee
Here the $P_{ni}$ are multivariate polynomials of order $n$
in $x_1$ and $x_2$ that are symmetric with respect
to the interchange of $x_1$, $x_2$, {\em and} 
$x_3\equiv 1-x_1-x_2$.  Because there can be
more than one polynomial of a given order, the
second subscript $i$ differentiates the
possibilities.  The sum on $n$ is truncated at a finite order $N$,
in order to have a finite number of equations; convergence
with respect to $N$ is investigated below. 

As shown in \cite{SymPolys}, the polynomials $P_{ni}$
can be constructed from linear combinations of
$C_{ml}(x_1,x_2)=C_2^m(x_1,x_2) C_3^l(x_1,x_2)$, where $C_2$
and $C_3$ are given by
\be \label{eq:basepolys}
C_2(x_1,x_2)=x_1^2+x_2^2+x_3^2, \;\;
C_3(x_1,x_2)=x_1 x_2 x_3,
\ee
and $2m+3l\leq n$.
For our purposes in this work, the particular
linear combinations are chosen to be orthonormal
with respect to the norm 
\be
\int_0^1 dx_1 \int_0^{1-x_1} dx_2\, x_1 x_2 x_3 P_{ni}(x_1,x_2) P_{mj}(x_1,x_2) =\delta_{nm}\delta_{ij}.
\ee
This norm has a different weight than used in \cite{SymPolys},
and the orthonormal polynomials are therefore slightly different.

Projection of the auxiliary equation (\ref{eq:LFCCaux}) onto the basis functions 
$\sqrt{y_1 y_2 y_3}P_{ni}(y_1,y_2)$ yields a matrix representation
\bea
\lefteqn{\sum_{mj} \left[(1+\Delta)A_{ni,mj}
   -3\left(1+\frac12\Delta\right)B_{ni,mj}+\frac32 g C_{ni,mj}\right]a_{mj}}&& \\
&& \rule{0.5in}{0mm} 
+\sum_{mj}\sum_{lk} \left[9g D_{ni,mj,lk}+\frac92 g F_{ni,mj,lk}\right]a_{mj} a_{lk}
+\frac{g}{6} G_{ni}=0,  \nonumber
\eea
with
\bea
A_{ni,mj}&\equiv & \int_0^1 dy_1 \int_0^{1-y_1} dy_2\, y_1 y_2 y_3 P_{ni}(y_1,y_2)
   P_{mj}(y_1,y_2)=\delta_{nm}\delta_{ij}, \\
B_{ni,mj} &\equiv & \int_0^1 dy_1 \int_0^{1-y_1} dy_2 \, y_2 y_3 P_{ni}(y_1,y_2) P_{mj}(y_1,y_2), \\
C_{ni,mj} &\equiv &\int_0^1 dy_1 \int_0^{1-y_1} dy_2 \, y_1 P_{ni}(y_1,y_2) 
                       \int_0^{1-y_1} dx_1 P_{mj}(y_1,x_1), \\
D_{ni,mj,lk}&\equiv & \int_0^1 dy_1 \int_0^{1-y_1} dy_2\, y_1 y_2 P_{ni}(y_1,y_2) \\
&& \rule{0.5in}{0mm} \times
\int_{y_1/(1-y_2)}^1 \frac{d\alpha_1}{\alpha_1} \int_0^{1-\alpha_1} d\alpha_2 
       P_{mj}(y_1/\alpha_1,y_2) P_{lk}(\alpha_1,\alpha_2), \nonumber \\
F_{ni,mj,lk}&\equiv & \int_0^1 dy_1 \int_0^{1-y_1} dy_2 \, y_1 y_2 P_{ni}(y_1,y_2) \\
&& \rule{0.5in}{0mm} \times
\int_{y_1+y_2}^1 \frac{d\alpha_1}{\alpha_1^2} \int_0^{1-\alpha_1} d\alpha_2
        P_{mj}(y_1/\alpha_1,y_2/\alpha_1)P_{lk}(\alpha_1,\alpha_2), \nonumber
\eea
and
\be
G_{ni}\equiv \int_0^1 dy_1 \int_0^{1-y_1} dy_2 P_{ni}(y_1,y_2).
\ee
The self-energy $\Delta$, defined in (\ref{eq:Delta}), is given by
\be
\Delta=g\sum_{ni}G_{ni}a_{ni}.
\ee
All of these integrals can be done analytically.  With {\em Mathematica},
the orthonormal polynomials can be generated and the integrals computed.
In practice, however, because of the multiple integrations and the 
large number of matrix elements, particularly for the rank-three matrices
$D$ and $F$, the calculation can be quite slow.  A more efficient approach
is to use Gauss--Legendre quadrature to compute the integrals; this can
be exact, if the quadrature is of sufficiently high order, because the
integrands of $B$ and $C$ are explicitly polynomial and those of
$D$ and $F$ can be transformed to be polynomial.

The transformation for $D$ is to first change variables from
$\alpha_1$ to $z_1=y_1/\alpha_1$.  With a change in order of
integration, this leaves
\bea
D_{ni,mj,lk}&\equiv & \int_0^1 \frac{dz_1}{z_1} \int_0^{z_1} dy_1 \int_0^{1-z_1} dy_2\, y_1 y_2 P_{ni}(y_1,y_2) \\
&& \rule{0.5in}{0mm} \times P_{mj}(z_1,y_2)
\int_0^{1-y_1/z_1} d\alpha_2  P_{lk}(y_1/z_1,\alpha_2). \nonumber
\eea
To complete the transformation, we change variables from $y_1$ to $z_2=y_1/z_1$.
This yields
\bea
D_{ni,mj,lk}&\equiv & \int_0^1 dz_1 \int_0^1 dz_2 \int_0^{1-z_1} dy_2\, z_1 z_2 y_2 P_{ni}(z_1z_2,y_2) \\
&& \rule{0.5in}{0mm} \times P_{mj}(z_1,y_2)
\int_0^{1-z_2} d\alpha_2  P_{lk}(z_2,\alpha_2). \nonumber
\eea
For $F$, we change the order of integration, to place the integral over $\alpha_1$ first,
and change variables from $y_1$ and $y_2$ to $z_1=y_1/\alpha_1$ and $z_2=y_2/\alpha_1$,
to obtain
\bea
F_{ni,mj,lk}&\equiv & \int_0^1 d\alpha_1 \int_0^1 dz_1 \int_0^{1-z_1} dz_2 \, 
    \alpha_1^2 z_1 z_2 P_{ni}(\alpha_1 z_1,\alpha_1 z_2) \\
&& \rule{0.5in}{0mm} \times P_{mj}(z_1,z_2)
    \int_0^{1-\alpha_1} d\alpha_2 P_{lk}(\alpha_1,\alpha_2). \nonumber
\eea

Once the matrix elements are calculated, we solve the nonlinear system
with the root finding procedure of {\em Mathematica}.
The initial guess for $a_{mj}$ is taken to be zero, so that the
solution found will correspond to the smallest contribution from 
$\tilde t_2$.

As a simple test of the polynomial basis, we can consider the restricted three-body
problem used in \cite{SymPolys}.  This is the simplification of the Fock-space
truncation that excludes two-two scattering.
The one-body equation remains the same as (\ref{eq:tildepsi1}),
but the three-body equation becomes
\be \label{eq:tildepsi3no22}
\frac16\frac{g}{\sqrt{x_1x_2x_3}}
   +\left(\frac{1}{x_1}+\frac{1}{x_2}+\frac{1}{x_3}-\frac{M^2}{\mu^2}\right)
     \tilde\psi_3(x_1,x_2,x_3)=0. 
\ee
This is easily rearranged to yield
\be
\tilde\psi_3=-\frac16\left(\frac{1}{x_1}+\frac{1}{x_2}+\frac{1}{x_3}-\frac{M^2}{\mu^2}\right)^{-1}
                      \frac{g}{\sqrt{x_1x_2x_3}},
\ee
and the one-body equation then reduces to a nonlinear equation for the mass
\be \label{eq:no22M2}
\frac{M^2}{\mu^2}=1
   -\frac{g^2}{6}\int\frac{dx_1 dx_2}{x_1 x_2 x_3}
       \left(\frac{1}{x_1}+\frac{1}{x_2}+\frac{1}{x_3}-\frac{M^2}{\mu^2}\right)^{-1}.
\ee
In \cite{SymPolys}, the equivalent of this equation was solved for $g$ as
a function of $M/\mu$, which is easily computed directly.  Here, in order to 
facilitate comparisons, we compute the mass ratio as a function of $g$, with
use of the root finding procedure of {\em Mathematica}.

\begin{figure}[ht]
\vspace{0.22in}
\centerline{\includegraphics[width=12cm]{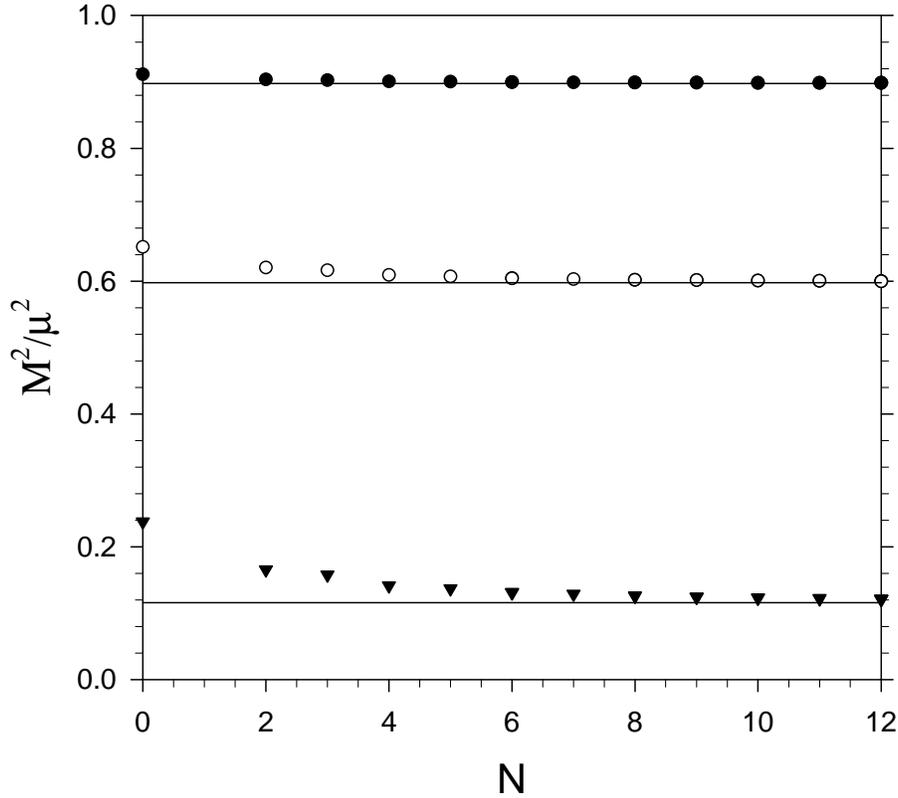}}
\caption{\label{fig:check} 
Comparison of the numerical and analytic solutions to the Fock-space
truncated equations without the two-two scattering contribution.
The mass-squared ratio $M^2/\mu^2$ is plotted against the highest
order $N$ of multivariate polynomial used, for three different values
of the coupling $g$.  The horizontal lines are the analytic results
for the chosen values of $g$, which are 0.5, 1, and 1.5.
}
\end{figure}

The two equations (\ref{eq:tildepsi1}) and (\ref{eq:tildepsi3no22}) can also
be solved by the same function expansion method as above.
The polynomial basis used here is different than that used in \cite{SymPolys},
due to a different choice of weighting factor for the inner product of
polynomials.  A comparison of the direct calculation of the mass squared
with the polynomial-basis calculation is shown in Fig.~\ref{fig:check}.
This demonstrates that the numerical calculation reproduces the analytic
result with rapid convergence, as previously observed in \cite{SymPolys}.
%

\begin{figure}[ht]
\vspace{0.2in}
\centerline{\includegraphics[width=12cm]{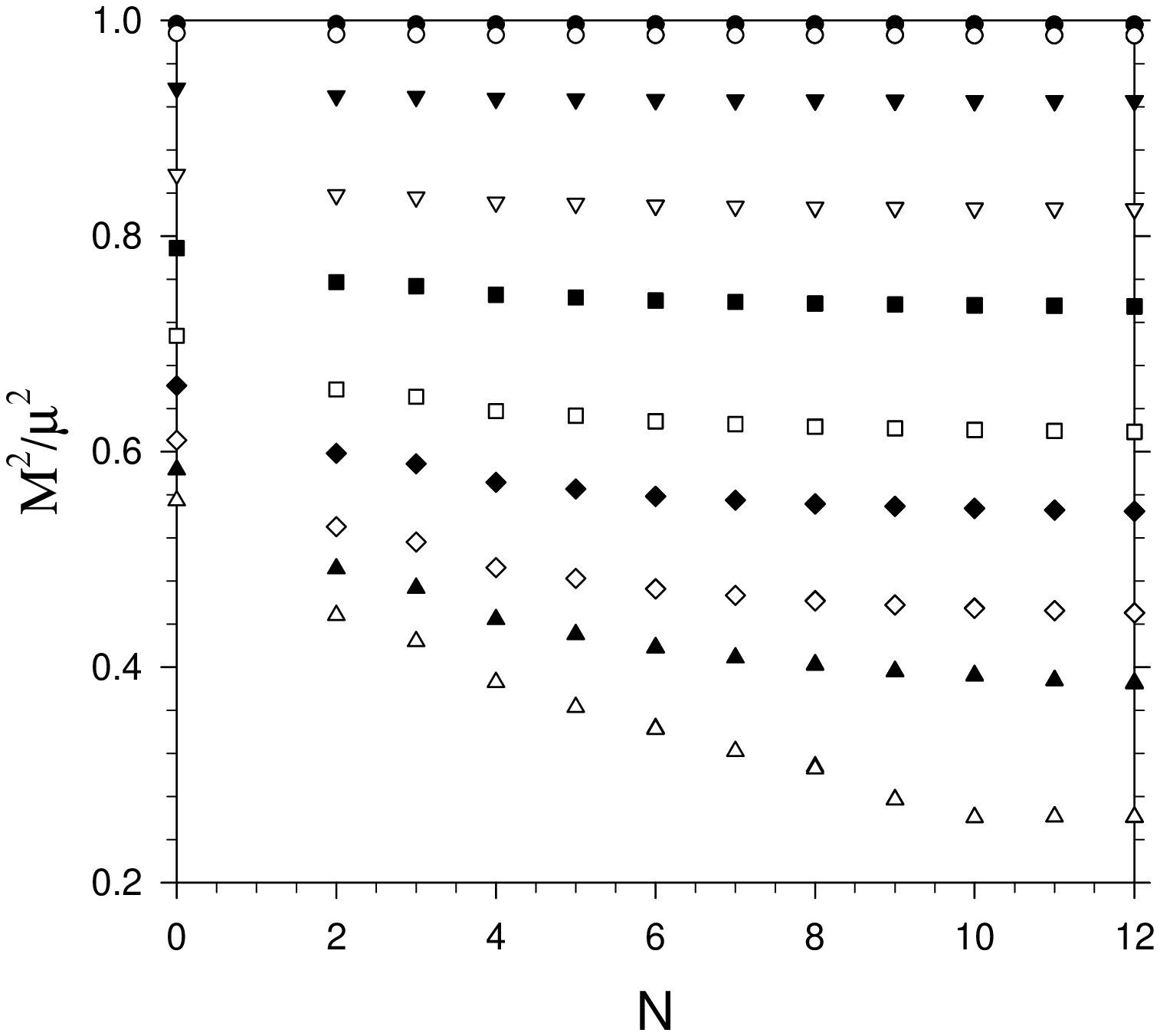}}
\caption{\label{fig:convergence} 
Convergence test for the numerical solution of the LFCC equations.
The mass-squared ratio $M^2/\mu^2$ is plotted versus the maximum
order included in the polynomial basis for the following
sequence of $g$ values: 0.1, 0.2, 0.5, 0.8, 1.0, 1.2, 1.3, 1.4, and 1.5.
}
\end{figure}

We now turn to the question of convergence for Eq.~(\ref{eq:LFCCaux}),
the equation for $\tilde t_2$.  Figure~\ref{fig:convergence} shows
the results for $M^2/\mu^2$, as the polynomial order $N$ is varied,
with $g$ assigned a broad range of values, from 0.1 to 1.5.
Although the convergence is not as rapid as that for the 
simpler test, the results do converge quite well.
%


\end{document}